\documentstyle[11pt,aaspp4]{article}

\begin{document}

\title{Radial Velocity Studies of Close Binary Stars.~II}

\author{Slavek M. Rucinski and Wenxian Lu\\
e-mail: {\it rucinski@astro.utoronto.ca, lu@astro.utoronto.ca}}

\affil{David Dunlap Observatory, University of Toronto \\
P.O.Box 360, Richmond Hill, Ontario L4C~4Y6, Canada}

\centerline{\today}

\begin{abstract}
Radial velocity measurements and simple sine-curve fits to the orbital 
velocity variations are presented for the second set of ten contact binary 
systems. Eight systems are of the A-type: AH Aur, CK Boo, DK Cyg, UZ Leo, 
XZ Leo, V839 Oph, GR Vir and NN Vir. V842 Her is the only W-type, while SV Equ 
appears to be a semi-detached system seen at a low orbital inclination rather 
than a contact binary. Several of the studied system are prime candidates 
for complete light and radial-velocity synthesis solutions.
\end{abstract}

\keywords{ stars: close binaries - stars: eclipsing binaries -- 
stars: variable stars}

\section{INTRODUCTION}
\label{sec1}

This paper is a continuation of the radial velocity studies of close 
binary stars started by Lu \& Rucinski \markcite{LR99} 
(1999 = Paper~I). The main 
motivation of this program have been determination of mean ($\gamma$) 
velocities for Hipparcos stars in order to determine spatial velocity 
vectors, with an important by-product of preliminary values of 
mass-ratio from simple sine-curve fits to the data. The program was started 
with contact binaries, mostly because of the evidence of their very high 
spatial frequency of occurrence relative to F--K dwarfs at the level of 
1/100 -- 1/80  (Rucinski \markcite{ruc98} 1998), 
but is slowly expanding to include 
other close binary systems accessible to the 1.8 meter class 
telescopes at medium spectral resolution of about R = 10,000 -- 15,000. 

The paper is structured in the same way as Paper~I in that it 
consists of two tables containing the radial velocity 
measurements (Table~\ref{tab1}) and 
their sine-curve solutions (Table~\ref{tab2}) 
and of brief summaries of previous studies for individual systems. 
The reader is referred to Paper~I for technical details of the 
program. In short, all observations described here were made with the 
1.88 meter telescope of the David 
Dunlap Observatory (DDO) of the University of Toronto. The 
Cassegrain spectrograph giving the scale of 
0.2 \AA/pixel, or about 12 km/s/pixel, was used; 
the pixel size of the CCD was 19~$\mu$m. A relatively wide 
spectrograph slit of 300~$\mu$m corresponded to the angular size on 
the sky of 1.8 arcsec and the projected width of 4 pixels.   
The spectra were centered at 5185 \AA\ with the spectral coverage of 
210 \AA. The exposure times were typically 10 minutes long, with the longest 
exposures for fainter systems not exceeding 15 minutes. 

The data in Table~\ref{tab1} are organized the same was as 
in Paper~I. Table~\ref{tab2} is slightly different from that in Paper~I as 
it now provides information about relation between the specroscopically
observed epoch of the primary-eclipse T$_0$ and the recent photometric 
determinations in the form of the (O--C) deviations for the number of 
elapsed periods E. It also contains, in the first column, below the star name,
our new spectral classifications of the program objects. The classification
spectra, typically two per object, were obtained using the same spectrograph, 
but with a different grating giving the dispersion of 0.62 \AA/pixel 
in the range 3850 -- 4450 \AA. Several spectral-classification standards were
observed and then the program-star spectra were ``interpolated'' between them
in terms of relative strenghts of lines 
known as reliable classification criteria.

In the radial-velocity solutions of the orbits, the data have  
been assigned weights on the basis of  our ability to resolve the 
components and to fit independent Gaussians to each of the broadening-function 
peak. Weight equal to zero in 
Table~\ref{tab1} means that an observation was not used 
in our orbital solutions; however, these observations may be utilized in 
detailed modeling of  broadening functions, if such are 
undertaken for the available material. The full-weight points are 
marked in the figures by filled symbols while half-weight ones are marked 
by open symbols. Phases of  the observations with zero weights are 
shown by short  markers in the lower parts of the figures; they were 
usually obtained close to the phases of orbital conjunctions.

Because our data had been collected usually within one or two consecutive 
observing seasons, the orbital solutions were done by fixing the values 
of the orbital period. The solutions for the four circular-orbit
parameters, $\gamma$, K$_1$, K$_2$ and T$_0$, 
were obtained in the following way:
First, two independent least-squares solutions for 
each star were made using the same programs as described in Paper~I. They 
provided preliminary values of the amplitude, $K_i$, of the
mean velocity $\gamma$s and the initial (primary eclipse)
epoch $T_0$. Then, one combined solution for both amplitudes and the common 
$\gamma$ was made with the fixed mean 
value of T$_0$. Next, differential corrections the K$_1$, K$_2$, and T$_0$ 
were determined, providing best values of the four parameters. These 
values are given in Table~\ref{tab2}. 
The corrections to $\gamma$, K$_1$, K$_2$, and T$_0$ were finally 
subject to a ``bootstrap'' process (several thousand solutions with randomly 
drawn data with repetitions) to provide the median values and ranges of the
parameters. As is common in the application of this method,
the bootstrap one-sigma ranges were found to be
systematically larger than the differential-correction, linear 
least-squares estimates so that we have 
adopted  them as measures of  uncertainty of parameters in Table~\ref{tab2}.

Throughout the paper, when the errors are not written otherwise, we 
use notation of the standard mean 
errors in terms of the last quoted digits, e.g. the number 0.349(29) 
should be interpreted as $0.349 \pm 0.029$. 

\placetable{tab1}

\placetable{tab2}

\section{RESULTS FOR INDIVIDUAL SYSTEMS}
\label{sec2}

\subsection{AH Aur}

The binary has been discovered by Tsesevich \markcite{tze54} (1954) 
and early light curves were published by Hinderer \markcite{hin60}
(1960). The system was never observed for a study of  radial velocity 
variations; even photometrically, it 
is one of the least observed contact binaries. It shows a light curve 
with amplitude slightly larger than 0.5 mag and with indications of partial 
eclipses. Although it was included in the Hipparcos observing list, it 
has not been included in the analysis by Rucinski \& Duerbeck 
\markcite{RD97} (1997 = RD97) 
because the large error in its parallax resulting in the absolute-magnitude 
error of 0.72 mag.  The $(B-V)$ color from the Tycho 
experiment 0.55(8) suggested a spectral type late F -- early G, 
which is in excellent agreement with our spectral type F7V; this is
also consistent with the absolute-magnitude calibration 
$M_V = M_V\,(\log P, B-V)$ of RD97 which givs $M_V = 3.1$. 

The most recent available photometric timing of the primary eclipse was that by 
Agerer \& Hubscher \markcite{AH96}(1996); 
this timing was used as the first guess for our radial-velocity solution. 
In our orbital solution, we assumed the 
orbital period following the 1985 edition of the
General Catalogue of Variable Stars, $P = 0.4942624$ days. 
The  $(O-C)$ deviation for the primary eclipse epoch T$_0$
is relatively large and equals 0.0206 days 
which is much larger than the error of determination of T$_0$. 
This shift may be partly due to an obvious 
asymmetry in the radial-velocity curve of the less-massive component 
in the first half of the orbital cycle (see Figure~\ref{fig1}).
If we fix the T$_0$ at the epoch given by Agerer \& Hubscher 
\markcite{AH96} (1996), 
the asymmetry remains and the fit is only slightly modified. The values 
of the orbital parameters are then: ($\gamma$, K$_1$, K$_2$) = (30.96, 47.36, 
279.68) km/s. The mass-ratio 
remains then at a relatively low value of $q = 0.17$,
but values of velocity amplitudes which determine the masses
are slightly changed. The system needs a modern photometric study.

\placefigure{fig1}

\subsection{CK Boo}

Bond \markcite{bon75} (1975) noted diffuse spectral lines and then obtained a 
fragmentary light curve indicating that the star is a W~UMa-type binary. 
Since then, the binary has been subject of several time-of-minima studies, 
the most recent one by Muyesseroglu, Gurol \& Selam \markcite{muy96}
(1996). We have taken the value of the period, $P=0.3551501$, from the 
study of Aslan \& Derman \markcite{AD86} (1986).

The light curves of the system were presented by Krzesinski, Mikolajewski, 
Pajdosz \& Zola \markcite{krz91} (1991) and 
Jia, Liu \& Huang \markcite{jia92} (1992). The light curves 
are relatively shallow indicating partial eclipses and, 
consequently, difficulties with light-curve-synthesis solutions. 
Krzesinski et al.\ attempted a solution 
which included a determination of the photometric mass ratio. 
They derived $q_{ph} = 0.52$ or 0.59 
(depending on the assumptions concerning spots) and found that the 
system is of the W-type, i.e.\ with the primary, deeper minimum corresponding 
to eclipses of the smaller, less massive star in the system.
The system was included in the Hipparcos study (RD97) with a relatively 
poor determination of the absolute magnitude, $M_V = 2.99(44)$,
which is however consistent with $(B-V)_0 = 0.54$ and a spectral type of F6V as 
estimated by Krzesinski et al.\ (1991). Our estimate of the spectral type
is slightly later, F7/8V.

The only radial velocity study that we are aware of is by Hrivnak 
\markcite{hri93} (1993). 
His preliminary spectroscopic determination of the mass-ratio, $q_{sp} = 0.16$, 
was entirely different from the photometric solution of Krzesinski et 
al.\ (1991). Also, the type of the system was found to be A, not W;
i.e.\ it is the more massive star which
is eclipsed at deeper minimum. Our data fully confirm that the system is of the 
A-type, but our mass-ratio is even smaller 
than that of Hrivnak, $q_{sp} =  0.11$, 
although the error of the spectroscopic determination is 
relatively large for such small values of $q$.

\subsection{DK Cyg}

This contact system has not been studied spectroscopically yet, 
although it has been a subject of numerous photometric investigations 
since the discovery by Guthnick \& Prager \markcite{GP27} (1927). 
The most recent photometry and light curve have been given by Awadalla 
\markcite{awa94} (1994); this study 
suggested a systematic period change. We adopted $P=0.470691$ days for our 
data, a number which is based on the values given by 
Awadalla \markcite{awa94} 1994) and Binnendijk \markcite{bin64} (1964). 

Because of the total eclipses at shallower minimum (a clear A-type), 
Mochnacki \& Doughty \markcite{MD72}
(1972) recognized that the system would be convenient 
one for their (one of the first) contact-model light-curve-synthesis solutions; 
they used the light curve of Binnendijk \markcite{bin64} 
(1964). Their photometric mass-ratio was 
$q_{ph}=0.33 \pm 0.02$. Our entirely independent spectroscopic result of 
$q_{sp} = 0.32 \pm 0.04$ agrees fully with 
this determination confirming that systems with total eclipses 
can give excellent photometric solutions, in 
contrast to systems with partial eclipses. Mochnacki and Doughty 
\markcite{MD72} (1972)  
pointed out that the spectral type of the system, judged by its color, 
is probably F0--F2. This prediction did not agree with the direct estimate of  
the spectral type at A6--8V given by Hill, Hilditch, Younger \& 
Fischer \markcite{hill75} (1975). We also estimated the spectral type
at A8V. The Stromgren color $b-y=0.24$ (Hilditch \& Hill \markcite{hild75} 
1975) suggests a spectral type A8 -- F0. The Hipparcos parallax is small and 
has a large error ($\epsilon M_V = 0.9$) 
so that the system was not included in RD97. 

The system was included in the list of near-contact binaries of Shaw 
\markcite{sha94} (1994), 
but as far as we can establish, it an excellent example of an A-type 
contact system without any particular complications so there are no reasons to
consider it a ``near-contact'' one.

\subsection{SV Equ}

This relatively long-period system (0.881 days) has been 
discovered by Catalano \& Rodono \markcite{CR66}
(1966) and then photometrically studied by Eggen \markcite{egg78}
(1978). Because of the color $b-y=0.145$ suggesting a spectral type 
of A5 and of the shallow (0.15 mag) light curve, Eggen considered the 
system to be a contact binary of little interest in the 
context of genuine early-type systems. The system has been somewhat 
forgotten except for its inclusion among near-contact binaries in the 
compilation of Shaw \markcite{sha94} (1994). The Hipparcos parallax is poor 
in this case giving $M_V = 3.7(9)$. 

Recent photometric observations of SV Equ were reported by Cook 
\markcite {coo97} (1997) 
who gave the new time-of-minimum prediction with the period $P=0.88097307$
days.  These observations were obtained very close in 
time to our observations, but they disagree in the time of minimum T$_0$. 
We do not see any obvious reasons why the 
$(O-C)$ for contemporaneous observations should be as large as $-0.028$ 
days, but note that the graph of the data in Cook (1997) indicates 
rather large photometric errors.

In spite of attempts to isolate lines of the secondary component with 
different template spectra, out spectroscopic observations led to detection 
of only one component. Thus, the star is not a contact binary 
(EW) as it was classified before, but most probably a semi-detached system (EB) 
seen at a low orbital inclination angle.

\subsection{V842 Her (NSV 7457)}

The variability was discovered by Geyer, Kippenhahn \& Stroheimer 
\markcite{gey65} (1965), 
but the real nature of the star as a contact binary was identified 
relatively recently by Vandenbroere \markcite{van93} (1993) and, apparently 
independently, by Nomen-Torres \& Garcia-Moreno \markcite{NG96}
(1996). We analyzed all the recent determinations of eclipse timings, 
as published by Vandenbroere \markcite{van93} (1993), 
Diethelm \markcite{die94} (1994), 
Nomen-Torres \& Garcia-Moreno \markcite{NG96} (1996)
and Agerer \& Huebscher \markcite{AH97} \markcite{AH98a}
(1997, 1998a), and found the following ephemeris:
JD(Hel) = 2,447,643.1771(29) + 0.41903201(56) E. As for the other systems,
only the period was used to determine the phases of the spectroscopic
observations. 

V842 Her has not been yet observed for radial velocity changes. There exist 
also no previous estimates of its spectral type or color. It is the only W-type 
system in the current study; coincidentally, it is also the only system without a 
Hipparcos parallax measurement in this group of ten contact systems. 
The orbital period of 0.42 days is somewhat 
long for a typical W-type system and our spectral
classification of F9V is unusually early for a W-type system. The 
light curve has a moderately large amplitude of about 0.6 
mag and the primary (deeper) eclipses appear to be total or very close
to total so that the system has a potential of an excellent
combined, light and radial-velocity solution.

\placefigure{fig2}

\subsection{UZ Leo}

Variability of this star was discovered by Kaho \markcite{kah37}
(1937), but the correct type was identified much later 
by Smith \markcite{smi54} (1954).  Since then, there have been 
many photometric studies of the system, but no radial-velocity 
studies. For guidance on the orbital phases, we used the recent determination 
of Agerer and Huebscher \markcite{AH98a}
(1998a). To phase our observations, we used the 
value of the period from the study of  Binnendijk \markcite{bin72}
(1972).

The modern light-curve synthesis solutions of Vinko, Hegedues \&
Hendry \markcite{vin96} (1996) arrived at two 
similar possible values of the mass-ratio: $q_{ph}=0.233$
 and 0.227. They differ rather substantially from our 
spectroscopic value, $q_{sp} = 0.303(24)$. The system is clearly of the 
A-type with a large-amplitude light curve and total eclipses offering 
excellent prospects for a combined radial-velocity/photometric solution. 
The Hipparcos parallax placed the system slightly outside the $M_V$ 
error threshold of 0.5 mag used in RD97. 
The resulting relatively faint absolute magnitude of $M_V = 3.75(55)$
agrees better with the spectral type of F2--3V, implied by $(B-V) = 0.373$, 
than with the direct estimate of A7V (Vinko et al.\ 1996). Our estimate
of the spectral type is A9/F1V.

\subsection{XZ Leo}

The system was discovered by Hoffmeister \markcite{hof34}
(1934). The recent time of minimum 
has been taken from Agerer \& Huebscher \markcite{AH97}
(1997) while the period is that 
determined by Niarchos, Hoffman \& Duerbeck \markcite{nia94} (1994). 

The system was never observed spectroscopically. Niarchos et al.\ (1994) 
made a plausible and apparently correct 
assumption that the system is of the A-type 
and attempted to determine the mass-ratio. Their value, $q_{ph} = 0.726$, 
is very far from our spectroscopic determination, $q_{sp} = 0.348(29)$, 
once again demonstrating the dangers of spectroscopically-unconstrained 
light-curve solutions for partially eclipsing 
systems. They attempted to estimate the spectral type and preferred the 
range A7 to F0 rather than the 
previous estimates of A5 to A7. The Tycho's experiment color 
$(B-V)_T = 0.45(7)$
indicates a mid-F spectral type. Our spectral type is A8/F0V so that the color
does not agree with the spectral type.

\subsection{V839 Oph}

Variability of the star was discovered by Rigollet \markcite{rig47}
(1947). The system has 
not been yet observed for radial-velocity variations. The recent timing of 
the minimum was by Agerer \& Huebscher \markcite{AH98b} (1998b) while the 
period used for phasing of our observations was determined by 
Akalin \& Derman \markcite{AD97} (1997). 

In spite of partial eclipses and light curve instabilities, the system was 
subject to many photometric contact-model solutions. The most recent one 
by Akalin \& Derman (1997) arrived at the photometric 
mass-ratio of $q_{ph} = 0.40$. This substantially differs from our 
spectroscopic determination of $q_{sp} = 0.305(24)$. 
The system is of the A-type.

V839 Oph was included in the study of the Hipparcos data (RD97) with a 
moderately accurate determination of $M_V = 3.08(38)$. This absolute 
magnitude is consistent with the Tycho color $(B-V)_T = 0.62$,
with $(b-y) = 0.41$ (Hilditch \& Hill 1975) and the spectral type 
of F8V (Akalin \& Derman 1997) under an assumption that the reddening 
is relatively large, $E_{B-V} = 0.09$ (RD97). Our spectral type is F7V.

\subsection{GR Vir}

Strohmeier, Knigge \& Ott \markcite{str65}
(1965) noticed light variations of the star, 
but an independent discovery of Harris \markcite{har79}
(1979) led to its identification 
as a close binary system. The assumed period as well as the recent 
timing of the eclipse come from the photometric study of Cereda, Misto, 
Niarchos \& Poretti \markcite{cer88} (1988). 
Since the system was not recently observed, the accumulated uncertainty 
in the period as well as a likely change in its length since the 
observations of Cereda et al.\ have led to a large difference between the 
spectroscopic and predicted values of T$_0$ of 0.2208 days. 
We handled the implied problem of relating our 
radial-velocity to the photometric data of Cereda et al.\ by
assuming that the system is of the A-type, as indicated by the fact
that the secondary (shallower) eclipses are apparently total. 

The mass-ratio of the system, $q = 0.12$, is one of the smallest known for 
contact systems. Because of the total eclipses and the availability of 
the spectroscopic mass-ratio, the system is an ideal candidate for a 
new light-curve synthesis solution. 

GR Vir is the third and last system in this series which was 
included in the Hipparcos analysis (RD97). 
The absolute magnitude $M_V = 4.17(14)$
was the best determined among the three determinations. 
Because of its brightness and the features mentioned above, the system 
deserves much attention. The colors $(b-y)=0.37$ (Olsen 1983) and $(B-V)=0.55$
(Cereda et al. 1988) suggest the spectral type F9 -- G0. Our spectral
classification is F7/8V.

\placefigure{fig3}

\subsection{NN Vir (HD 125488)}

This is one of the stars whose variability has been detected by the 
Hipparcos satellite. Gomez-Ferrellad \& Garcia-Melendo \markcite{GG97}
(1997) correctly 
identified the type of variability and determined the period and the 
initial epoch T$_0$. The radial velocity variations have been observed 
by us for the first time. In spite of its large apparent brightness ($V=7.6$), 
it was excluded from the radial-velocity survey of early F-type 
stars by Nordstroem, Stefanik, Latham \& Andersen 
\markcite{nor97} (1997) because of 
the strong broadening of the spectral lines indicating rapid rotation 
and/or close binary character of the star. 

NN Vir has a relatively large mass-ratio, $q_{sp} = 0.491(11)$, 
which is rather infrequent among the A-type contact systems. 
The light curve of Gomez-Ferrellad \& Garcia-Melendo has a moderately 
large amplitude so that the system should be subject of a combined 
radial-velocity/photometric solution. 

NN Vir was not included in RD97, but its parallax is moderately large 
leading to relatively secure determination of the absolute magnitude 
of $M_V = 2.52(26)$. The color $b-y=0.25$ (Olsen \markcite{ols83} 1983) 
suggests the spectral type F3. This is confirmed by our direct
classification F0/1V.

\section{SUMMARY}

The paper brings radial velocity data for the second group of ten 
close binary systems that we observed, 
this time all observed at David Dunlap Observatory. All but SV Equ, 
which is probably a short-period semi-detached system, are contact binaries 
with both components clearly detected. Although we do not give the calculated
values of $(M_1+M_2) \sin^3i = 1.0385 \times 10^{-7}\,(K_1+K_2)^3\,
P({\rm day})\,M_\odot$, we note that for 
all 9 contact binaries they are relatively large and range 
between $1.17\,M_\odot$
(CK Boo) and $2.25\,M_\odot$ (UZ Leo) indicating that the orbital
inclination angles for all of them are not far from 90 degrees and
that the final, combined 
solutions of the light and radial-velocity variations should give reliable 
values of the masses.  

Although our observational selection for this group of ten systems
had been purely random, we found  -- after exclusion of SV~Equ -- 
that 8 systems among 9 W~UMa-type systems 
are of the A-type; only V842 Her is of the W-type.  A 
combination of factors could lead to this unusual 
preference over the W-type systems: (1)~Although it is generally 
more difficult to detect low-mass secondaries in A-type systems than in W-type 
systems because of the more extreme mass-ratios, our
instrumental setup is ideal for observations of the type presented here;
(2)~The A-type systems are, 
on the average, hotter and brighter so that they would be preferentially 
selected close to the faint limit of a magnitude-limited survey like ours; 
(3)~It is possible that in cases when spectral lines of 
secondary components had not been detected during first attempts by 
previous observers, the A-type systems were preferentially discarded in 
favor of the W-type systems since the latter would normally give full,
two-component orbital solutions. 

We point out which systems will be of greatest interests in  
the individual descriptions, in Section~\ref{sec2}. 
Here we note that three of our 
systems have very small mass-ratios: 0.11, 0.12 and 0.17 for 
CK Boo, GR Vir and AH Aur. Small values are common among the A-type systems, 
but the physical state of such extreme mass-ratio systems remains elusive. 
We note that the only W-type system, V842 Her has a relatively small 
(for a W-type system) mass-ratio of 0.26, while NN Vir has a relatively large 
mass-ratio of 0.49 (for an A-type system).

\acknowledgements

The research has made use of the SIMBAD database, operated at the CDS, 
Strasbourg, France and accessible through the Canadian Astronomy Data Centre, 
which is operated by the Herzberg Institute of 
Astrophysics, National Research Council of Canada.

\clearpage

\noindent
Captions to figures:

\bigskip

\figcaption[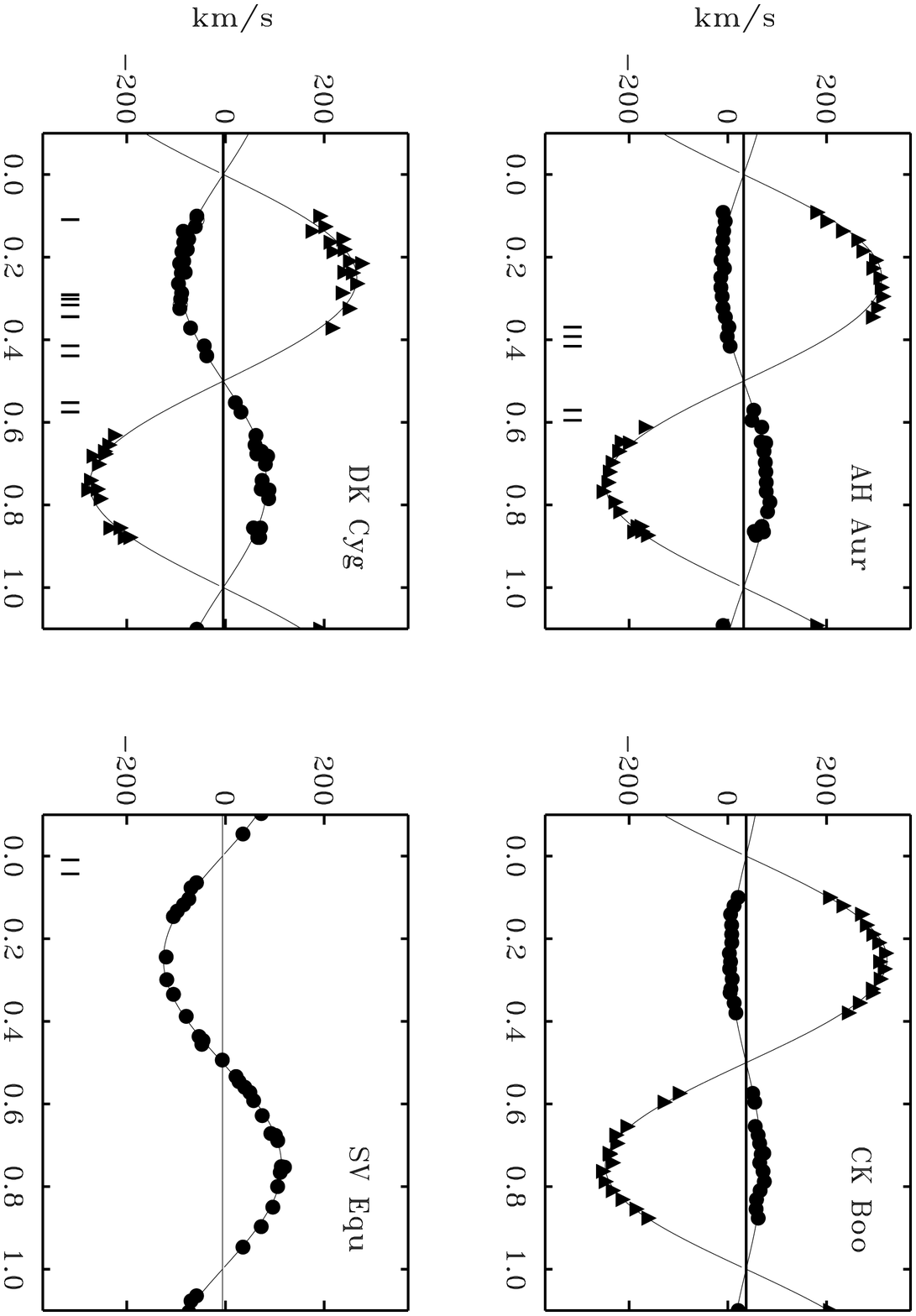] {\label{fig1}
Radial velocities of the systems AH Aur, CK Boo, DK Cyg and SV Equ,
are plotted in individual panels versus orbital phases. The thin lines 
give the respective circular-orbit 
(sine-curve) fits to the radial velocities.
SV Equ, previously considered a contact system, 
is apparently a semi-detached
(EB) binary. The remaining three systems are of the A-type, i.e.\
with eclipses of larger, more massive components at primary (deeper) 
minima. Short marks in the lower 
parts of the panels show phases of available 
observations which were not used in the solutions because of the 
blending of lines.
}

\figcaption[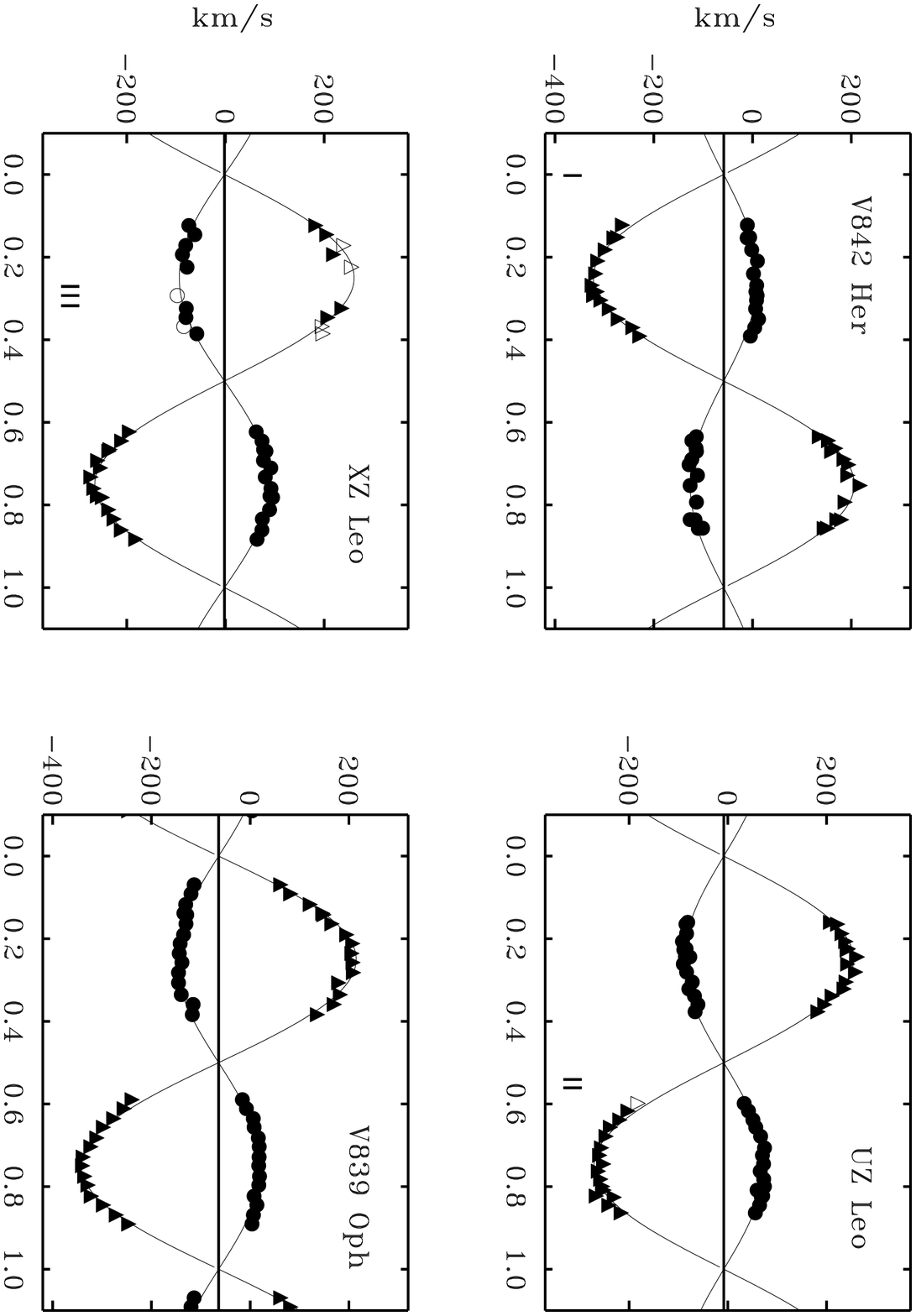] {\label{fig2}
Radial velocities of the systems 
V842 Her, UZ Leo, XZ Leo and V839 Oph,
in the same format as in Figure~1. V842 Her is the only W-type system
among the ten W~UMa-type systems described in this paper. 
Open symbols in this and the next figure
indicate observations given half-weights in the solutions. 
}

\figcaption[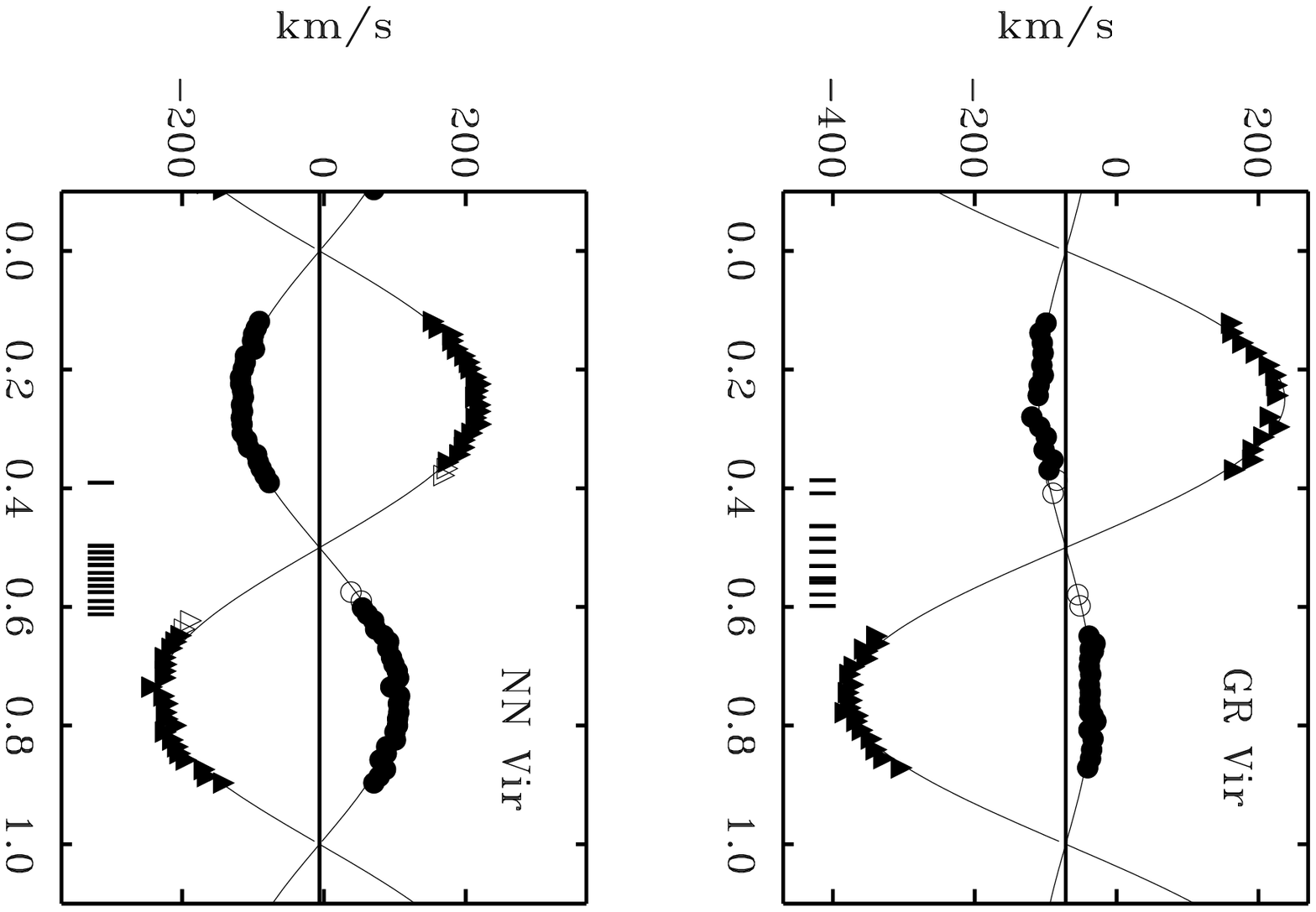] {\label{fig3}
Radial velocities of the systems GR Vir and NN Vir,
in the same format as in Figures~1 and 2. Because preliminary epochs of
eclipses were not known for these systems, many unneeded 
observations were taken at
conjuctions.
}

\clearpage

\begin{table}                 
\dummytable \label{tab1}      
\end{table}

\begin{table}                 
\dummytable \label{tab2}      
\end{table}

\end{document}


\pagestyle{empty}
 \begin{deluxetable}{lclrrrccc}
\tablecolumns{9}
\tablewidth{0pt}
\tablenum{2}
\tablecaption{Spectroscopic orbital elements of the second
ten of close binary systems}
\tablehead{
   \colhead{Name} &                
   \colhead{Type} &                
   \colhead{Other} &               
   \colhead{$\gamma$} &            
   \colhead{K$_1$} &               
   \colhead{$\epsilon_1$} &        
   \colhead{T$_0$ -- 2~400~000} &  
   \colhead{P (days)} &            
   \colhead{q =} \nl               
   \colhead{Sp.\ type}     &       
   \colhead{}               &      
   \colhead{names} &               
   \colhead{} &                    
   \colhead{K$_2$} &               
   \colhead{$\epsilon_2$} &        
   \colhead{(O--C) and [E]} &      
   \colhead{assumed} &             
   \colhead{$m_2/m_1$}             
}
\startdata
AH Aur & A & HD 256902      & 31.95(1.45) & 47.20(1.16)  & 
 5.42  & 50469.8591(16) & 0.494262    & 0.169(59)    \nl
 ~~~F7V& & HIP 30618      &             & 279.61(2.80) &
 15.15 & +0.0206 [1478] &             &              \nl
CK Boo & A & HD 128141      & 37.04(0.75) & 31.66(0.78)  &
 3.54  & 50539.9229(5)  & 0.355150    & 0.111(52)    \nl
 ~~~F7/8V &  & HIP 71319   &             & 285.31(1.63) &
  7.01 & +0.0056 [787]  &             &              \nl
DK Cyg & A  & BD+33 4304     & $-5.53(1.48)$ & 87.89(1.48)  &
  6.59 & 50331.2481(20) & 0.470691    & 0.325(39)    \nl
 ~~~A8V & & HIP 106574    &             & 270.46(3.36) &
 16.74 & +0.0049 [877]  &             &              \nl
SV Equ & EB & HIP 103431     & $-6.02(0.88)$ & 119.56(1.11) &
 4.23  & 50324.0880(15) & 0.880972    &              \nl
~~~A7V & &                &             &              &
       & $-0.0281$ [$-38$]  &             &              \nl
V842 Her & W & NSV 7457     & $-57.98(1.55)$ & 265.31(1.58) &
 12.88 & 50595.6856(8)  & 0.419032    & 3.852(24)    \nl
 ~~~F9V & & BD+50$^\circ$2255 &          & 68.88(1.60) &
  7.08 & $+0.0014$ [189]  &             & [=1/0.260]    \nl
UZ Leo &  A & BD+14$^\circ$2280 & $-8.18(1.12)$ & 79.69(1.12) &
  5.44 & 50595.6856(16) & 0.618043    & 0.303(24)    \nl
~~~A9/F1V & & HIP 52249      &             & 262.87(1.89) &
 9.93  & $-0.0006$ [214]  &             &              \nl
XZ Leo & A & BD+17$^\circ$2165 & $-2.03(1.43)$ & 91.49(1.80) &
 6.27  & 50520.6869(9)  & 0.487737    & 0.348(29)    \nl
~~~A8/F0V & & HIP 49204      &             & 262.55(2.60) &
 10.05 & $-0.0037$ [708]  &             &              \nl
V839 Oph & A& HIP 88946    & $-63.81(1.09)$ & 84.82(1.33) &
 6.52  & 50610.8436(9)  & 0.409005    & 0.305(24)    \nl
~~~F7V & &                &             & 278.52(2.01) &
 10.92 & $-0.0024$ [$-65$]  &             &              \nl
GR Vir & A& HD 129903      & $-71.72(0.89)$ & 37.78(1.00) &
 4.78  & 50541.9582(8)  & 0.346979    & 0.122(44)    \nl
~~~F7/8V & & HIP 72138      &             & 308.81(1.96) &
 12.10 & +0.2208 [14053] &            &              \nl
NN Vir & A& HD 125488      & $-6.24(0.65)$ & 112.68(0.71) &
 4.40  & 50595.3452(6)  & 0.480690    & 0.491(11)    \nl
~~~F0/1V& & HIP 70020      &             & 229.28(1.24) &
 7.84  & +0.0018 [145]  &             &              \nl
\tablecomments{The convention of naming the
components is that the subscript 1 designates
the component which is eclipsed at the
deeper minimum and is therefore the hotter one. 
The standard errors of the circular 
solutions in the table are expressed in units of last decimal places 
quoted; they are given in parantheses after each value. 
For example, the last table entry 0.491(11) should be 
interpretted as $0.491 \pm 0.011$. 
The average radial velocities ($\gamma$), the 
velocity amplitudes (K$_i$) and the standard unit-weight 
errors of the solutions ($\epsilon$) are expressed 
in $km\,s^{-1}$. The calculated moments of primary minima
are given by $T_0$ while their (O--C) deviations (in days) 
have been calculated from the most 
recent available ephemerides, as given in the text,
using the assumed periods and the number of epochs given by [E].}
\enddata
\end{deluxetable}